\documentclass[nofootinbib,superscriptaddress]{revtex4}
\usepackage{amsmath}
\usepackage{graphicx}

\begin{document}

\newcommand{\LF} {{\cal L}^{\rm F}}
\newcommand{\GSM} { {\cal G}^{\rm SM}}
\newcommand{\GSME} { {\cal G}^{\rm SM}_{\rm exten}}
\newcommand{\eg} {{\it e.g}}
\newcommand{\ie}  {{\it i.e}}
\newcommand{\et} {{\it et. al}}
\newcommand{\ba}{\begin{eqnarray}}
\newcommand{\ea}{\end{eqnarray}}
\newcommand{\bi}{\begin{itemize}}
\newcommand{\ei}{\end{itemize}}
\newcommand{\nn}{\nonumber}
\newcommand{\MeV}{{\rm \, MeV}}
\newcommand{\GeV}{{\rm \, GeV}}
\newcommand{\Or}{{\cal O}}
\def\Ord#1{\Or\left(#1\right)}
\def\Eq#1{Eq.~(\ref{#1})}
\newcommand{\ddbar}{D^0-\overline{D}{}^0}
\newcommand{\diag}{{\rm diag}}
\newcommand{\VKM}{V^{\rm CKM}}
\newcommand{\delKM}{\delta^{\rm KM}}
\newcommand{\CSM}{C^{\rm SM}}
\newcommand{\tr}{{\rm tr}}
\newcommand{\gW}{{g_2^\pm}}
\newcommand{\LMFV}{{\Lambda_{\rm MFV}}}
\newcommand{\LMFVF}{{\cal L}^{\Delta F=2}_{\rm MFV}}
\newcommand{\bp}{\begin{pmatrix}}
\newcommand{\ep}{\end{pmatrix}}
\newcommand{\Obsn}{{O_{b\to s\nu\bar \nu}}}
\newcommand{\Obdn}{{O_{b\to d\nu\bar \nu}}}


\title{\huge\vspace*{-.1cm}Brief Introduction to Flavor Physics\vspace*{.1cm}}

\author{{\Large Gilad Perez\vspace*{.1cm} }}

\email{gilad.perez@weizmann.ac.il  }
\affiliation{\large Weizmann Institute of Science \vspace*{1cm}}

\begin{abstract}
\centerline{\bf abstract \vspace*{.43cm}}
We consider the standard model (SM) quark flavor sector. We study its
structure in a spurionic, symmetry oriented approach. The SM
picture of flavor and CP violation is now experimentally verified,  
hence strong bounds on beyond the SM flavor structure follow.
We show how to parametrically derive such bounds, in a model independent manner, via minimal flavor violation
power counting. This min-review summarizes lectures given at the  ISSCSMB '08 international school.
It aims to give basic tools to understand how flavor and CP violation occur in the SM and its extensions.
It should be particularly useful for non-expert students who have mastered other aspects of the SM dynamics.
\vspace*{1cm}
  
\end{abstract}
\maketitle

\section{Introduction}

Flavors are just replication of states with identical quantum numbers.
The standard model (SM) consists of three such replications of the five fermionic representations of the SM
gauge group. Flavor physics describes the non-trivial spectrum and interactions of the flavor sector. 
What makes flavor physics particularly interesting is that the SM flavor sector is rather unique, its special characteristic makes it testable and predictive.
\footnote{Due to time limitation, this set of lectures discusses the quark sector only. Most of the concepts that are explained here can be directly applied to the lepton sector.} 
Let us list few of the SM unique flavor predictions:
\bi
\item Contains a single CP violating parameter.\footnote{The SM contains an additional flavor diagonal CP violating parameter, the strong CP phase, however, experimental data constrains it to be smaller than $\Or\left(10^{-10}\right)$, hence negligibly small.}
\item Flavor conversion is driven by three mixing angles.
\item To leading order, flavor conversion proceeds through weak charged current interactions.
\item To leading order, flavor conversion involves left handed (LH) currents.
\item CP violating processes must involve all the three generations.
\item The dominant flavor breaking is due to  the top Yukawa coupling, hence the SM posses a large approximate~
 global flavor symmetry [as shown below, technically it is given by $U(2)_Q\times U(2)_U\times U(1)_t \times U(3)_D$].
\ei
In the last four decades, or so, a huge effort was invested towards testing the SM predictions related to its flavor sector.
Recently, due to the success
of the B factories, the field of flavor physics has made a dramatic progress, culminated
in Kobayashi and Maskawa winning the Nobel prize.
It is now established that the SM contributions drive the observed flavor and CP violation (CPV)
in nature, via the Cabibbo-Kobayashi-Maskawa~\cite{Cabibbo:1963yz,Kobayashi:1973fv} description.
To verify that this is indeed the case, one can allow new physics (NP) to contribute 
to various clean observables which can be calculated precisely within the SM.
Analyses of the data before and after the B factories data have matured~\cite{Ligeti, NMFV,Buras:2009us}
demonstrate that the NP contributions to these theoretically clean processes cannot be bigger than $\Ord{30\%}$ 
of the SM contributions~\cite{UTFit,CKMFitter}.

Very recently, the SM passed another non-trivial test. 
The neutral $D$-meson system is the only one among the four neutral
meson systems ($K,D,B,B_s$) that is made of up-type quarks~(for formalism see~\eg~\cite{DDbarform} and Refs. therein). But this
is not the only unique aspect of this system:
(i) It is the only system where long distance contributions to
the mixing are orders of magnitude above the SM short
distance ones~\cite{Dlong}.
(ii) It is the only system where the SM contribution
to the CP violation in the mixing amplitude is expected to be below
the permil level~\cite{DCPV}.
The first point means that it is extremely difficult to theoretically
predict the width and
mass-splitting. The second point implies that, in spite of this
inherent uncertainty, $\ddbar$ mixing can unambiguously signal new
physics if CPV is observed. Present data~\cite{Ciuchini:2007cw,Gedalia:2009kh,otherD,combine,indirect}
implies that generic CPV contributions can be only of $\Ord{20\%}$ 
of the total (un-calculable) contributions to the mixing amplitudes, again consistent with the SM null prediction.

We have just given rather solid arguments for the validity of the SM flavor description. 
What else is there to say then? Could this be the end of the story?
We have several important reasons to think that flavor physics will continue to play a significant 
role in our understanding of microscopical physics at and beyond the reach of current colliders.
Let us first mention few examples that demonstrate the role that flavor precision tests played in the past:
\bi
\item The smallness of $\frac{\Gamma(K_L\to\mu^+\mu^-)}
  {\Gamma(K^+\to\mu^+\nu)}$ led to predicting a fourth (the charm)
  quark via the discovery of the  GIM mechanism~\cite{GIM};
\item The size of $\Delta m_K$ led to a successful prediction of the
  charm mass~\cite{Charm};
\item The size of $\Delta m_B$ led to a successful prediction of the
  top mass (for a review see~\cite{Franzini:1988fs} and Refs. therein).
\ei
This partial list demonstrates the power of flavor precision tests in term
of being sensitive to short distance dynamics.
Even in view of the SM successful flavor story, it is likely that there are missing experimental and theoretical ingredients as follows:
\bi
\item Within the
  SM, as mentioned, there is a single CP violating parameter. We shall see that the unique structure of the SM 
  flavor sector implies that CP violating phenomena are highly suppressed.
  Baryogenesis which requires sizable CP violating source therefore cannot be accounted for
  by the SM CKM phase.
   Measurements of CPV in flavor changing
  processes might provide evidences for additional sources coming from short distance physics.
\item The SM flavor parameters are hierarchical and most of them are small (excluding the top Yukawa and the CKM phase) which is denoted as the flavor puzzle.
This peculiarity might stem from unknown flavor dynamics. Though it might be related 
to a very short distance physics we can still get indirect information about its nature via combinations of flavor precision and high $p_T$ measurements.
 \item The SM fine-tuning problem, which is related to the quadratic divergence 
of the Higgs mass, generically requires new physics at, or below, the
  TeV scale. If such new physics have a generic flavor structure, it
  would contribute to flavor changing neutral current (FCNC) processes
  orders of magnitude above the observed rates. Or putting it differently, the flavor scale where NP is allowed to have a generic flavor structure is required to be larger than $\Ord{10^5}\,$TeV to be consistent with flavor precision tests, well above the electroweak symmetry breaking scale.
This implies an "intermediate" hierarchy puzzle ({\it cf.} the little hierarchy~\cite{LHierarchy} problem).
We use the term puzzle and not problem since in general the smallness of the flavor parameters, even within NP models,
implies the presence of approximate symmetries.
One can imagine, for instance, a situation where the suppression of the NP contributions to 
FCNC processes is linked with the SM small mixing angles and small the light quark Yukawas~\cite{NMFV}. In such a case this "intermediate" hierarchy is resolved in a technically natural way,
or radiatively stable manner and no fine tuning is required.\footnote{Unlike, say, the case of the $S$ electroweak parameter where in general one cannot associate an approximate symmetry with the limit of small NP contributions to $S$.}
\ei

\section{The standard model flavor sector}
The SM  quarks furnish three representations of the SM gauge group, $SU(3)\times SU(2)\times U(1)$:
$Q(3,2)_{1\over6}\times U(3,1)_{2\over3}\times U(3,1)_{-{1\over3}}$, where $Q,U,D$ stand for $SU(2)$ weak doublet, up type and down type singlet
quarks respectively.
Flavor physics is related to the fact that the SM consists of three replications/generations/flavors of these three representations.
The flavor sector of the SM is described via the following part of the SM Lagrangian
\begin{equation}
\LF=\overline{q^i} D \hspace*{-.25cm\slash}\ q^j \delta_{ij}+(Y_U)_{ij}\overline{ Q^i} U^j H_u+
(Y_D)_{ij}\overline{ Q^i} D^j H_d\,,\label{Lflavor}
\end{equation}
where  $D \hspace*{-.25cm\slash}\equiv D_\mu\gamma^\mu$ with $D_\mu$ being a covariant derivative, $q=Q,U,D$, within the SM with a single Higgs $H_u=i\sigma_2 H_d^*$ (however, the reader should keep in mind that at present, the nature and content of the SM Higgs sector is unknown) and $i,j=1,2,3$ are flavor indices.

If we switch off the Yukawa interactions the SM posses a large global, flavor symmetry,
$ \GSM$,\footnote{At the quantum level a linear combination of the diagonal $U(1)$'s inside the $U(3)$'s which corresponds to the axial current is anomalous.}
\begin{eqnarray}
  \label{GSM}
  \GSM= U(3)_Q\times U(3)_U\times U(3)_D\,.
\end{eqnarray}
Inspecting Eq. (\ref{Lflavor}) shows that the only non trivial flavor dependence in the Lagrangian is in the form of  Yukawa interactions. It is encoded in
a pair of $3\times 3$ complex matrices, $Y_{U,D}$.

\subsection{The SM quark flavor parameters}
Naively one might think that the number of the SM flavor parameters is given by $2\times 9=18$ real numbers
and $2\times 9=18$ imaginary ones, the elements of $Y_{U,D}$.
However, some of the parameters which appear in the Yukawa matrices are unphysical.
A simple way to see that (see \eg~\cite{Nirrev1} and Refs. therein) is to use the fact that a flavor basis transformation,
\begin{eqnarray}
  \label{Qtrans}
  Q \to V_Q Q\,, \qquad U\to V_U U\,, \qquad  D \to V_D  D\,,
\end{eqnarray}
leaves the SM Lagrangian invariant apart from redefinition of the Yukawas, 
 \begin{eqnarray}
  \label{Ytrans}
  Y_U \to V_Q Y_U V_U^\dagger\,, \qquad  Y_D \to V_Q Y_D V_D^\dagger\,,
\end{eqnarray}
where $V_i$ is a $3\times 3$ unitary rotation matrix.
Each of the three rotation matrices $V_{Q,U,D}$ contain three real parameters and six imaginary ones [the former ones correspond to the three generators of the $SO(3)$ group and the latter correspond to the rest, six generators, of the $U(3)$ group].
We know, however, that physical observables do not depend on our choice of basis.
Hence, we can use these flavor rotations to eliminate unphysical flavor parameters from $Y_{U,D}$.
Out of the 18 real parameters we can remove 9 ($3\times 3$) ones. Out of the 18 imaginary parameters we can remove
17  (3$\times6-1$) ones. We cannot remove all the imaginary parameters due to the
fact that the SM Lagrangian conserves a $U(1)_B$ symmetry.\footnote{More precisely only the combination $U(1)_{B-L}$ is non-anomalous.} Thus, there is a linear combination of the diagonal generators of $\GSM$ which is unbroken even in the presence of the Yukawa matrices and hence cannot be used in order to remove the extra imaginary parameter.

f
An explicit calculation shows that the 9 real parameters correspond to 6 masses and 3 CKM mixing angles,
while the imaginary parameter corresponds to the CKM celebrated phase.
To see that, we can define a mass basis where $Y_{U,D}$ are both diagonal.
This can be achieved by applying a bi-unitary transformation on each of the Yukawas:
\begin{eqnarray}
  \label{Qmtrans}
  Q^{u,d} \to V_{Q^{u,d}} Q^{u,d}\,, \qquad U\to V_U U\,, \qquad  D \to V_D  D\,,
\end{eqnarray}
which leaves the SM Lagrangian invariant apart from redefinition of the Yukawas, 
 \begin{eqnarray}
  \label{Ymtrans}
  Y_U \to V_{Q^u} Y_U V_U^\dagger\,, \qquad  Y_D \to V_{Q^d} Y_D V_D^\dagger\,,
\end{eqnarray}
the difference between the transformations used in Eqs.~(\ref{Qtrans},\ref{Ytrans})
and the ones above~(\ref{Qmtrans},\ref{Ymtrans}) is in the fact that each component of the $SU(2)$ weak doublets 
(denoted as $Q^u\equiv U_L$ and $Q^d\equiv D_L$) transforms independently. This manifestly breaks the $SU(2)$ gauge invariance, hence,
such a transformation makes sense only for a theory in which the electroweak symmetry is broken.
This is precisely the case for the SM where the masses are induced by spontaneous electroweak symmetry breaking 
via the Higgs mechanism.
Applying the above transformation amounts to "moving" to the mass basis. The SM flavor Lagrangian, in the mass basis, is given by (in a unitary gauge), 
\ba
\LF_m=\left(\overline{q^i} D \hspace*{-.25cm\slash}\ q^j \delta_{ij}\right)_{\rm NC}
+\begin{pmatrix} \overline{u_L} \,\overline{c_L}\, \overline{t_L}\end{pmatrix}
\begin{pmatrix}y_u&0&0\cr 0&y_c&0\cr 0&0&y_t\end{pmatrix} \begin{pmatrix} u_R \cr c_R \cr t_R \end{pmatrix} \left( v+h\right)+(u, c, t)\leftrightarrow(d,s, b)+{g_2\over\sqrt2}{\overline {u_{Li}}}\gamma^\mu
\VKM_{ij}d_{Lj} W_\mu^++{\rm h.c.}
,\,\label{Lflavormass}
\ea
where the subscript NC stands for neutral current interaction for the gluons, the photon and the $Z$ gauge bosons, $W^\pm$ stands for the charged electroweak gauge bosons, $h$ is the physical Higgs field, $v\sim 176\,$GeV, $m_i=y_i v$ and 
 $\VKM$ is the CKM matrix 
\ba\label{VCKM}
\VKM=V_{Q^u}V_{Q^d}^\dagger\,.
\ea
In general the  CKM is a $3\times3$ unitary matrix, with 6 imaginary parameters. However, as evident from Eq. (\ref{Lflavormass}), the charged current interactions 
are the only terms which are not invariant under individual quark vectorial $U(1)^6$ field redifinitons, 
\ba
u_i,d_j\to e^{i\theta_{u_i,d_j}}\,.
\ea 
The diagonal part of this transformation corresponds to the classically conserved baryon current while the non-diagonal, $U(1)^5$, part of the
transformation can be used to remove 5 out of the 6 phases, leaving the CKM matrix with a single physical phase.
Notice also that a possible permutation ambiguity for ordering the CKM entries is removed
since we have ordered the fields in \Eq{Lflavormass} according to their masses, light fields first. 
This exercise of explicitly identifying the mass basis rotation is quite instructive, and we have already learned several 
important issues regarding how flavor is broken within the SM (we shall derive the same conclusions using a spurion analysis in a symmetry oriented manner below in section~\ref{spurion}).
\bi
\item Flavor conversions only proceed via the three CKM mixing angles.
\item Flavor conversion is mediated via the charged current electroweak interactions. 
\item The charge current interactions only involve LH fields.
\ei

Even after removing all the unphysical parameters there are various possible forms for the CKM matrix.
For example, a parameterization used by the 
particle data group (PDG)~\cite{PDG}, is given by 
\ba\label{stapar}
\VKM=\begin{pmatrix}
c_{12}c_{13}&s_{12}c_{13}&
s_{13}e^{-i\delKM}\cr 
-s_{12}c_{23}-c_{12}s_{23}s_{13}e^{i\delKM}&
c_{12}c_{23}-s_{12}s_{23}s_{13}e^{i\delKM}&s_{23}c_{13}\cr
s_{12}s_{23}-c_{12}c_{23}s_{13}e^{i\delKM}&
-c_{12}s_{23}-s_{12}c_{23}s_{13}e^{i\delKM}&c_{23}c_{13}\cr \end{pmatrix},
\ea
where $c_{ij}\equiv\cos\theta_{ij}$ and $s_{ij}\equiv\sin\theta_{ij}$. The 
three $\sin\theta_{ij}$ are the three real mixing parameters while $\delKM$ is 
the Kobayashi-Maskawa phase. 

\subsection{CP violation}
The SM predictive power picks up once CPV is considered.
We have already proven that the SM flavor sector consists of a single CP violating parameter.
Once presented with a SM Lagrangian where the Yukawa matrices are given in a generic basis, it is not-trivial to determine whether CP is violated or not. This is even more challenging when discussing beyond the SM
dynamics where new CP violating sources might be present. 
A brute force way, to establish that CP is violated would be to show that no field redefinitions
would render a real Lagrangian.
For example consider a Lagrangian with a single Yukawa matrix,
\ba\label{Yukpairs}
{\cal L}^Y=Y_{ij}\overline{\psi^i_{L}}\phi\psi^j_{R}
+Y_{ij}^*\overline{\psi^j_{R}}\phi^\dagger\psi^i_{L},
\ea
where $\phi$ is a scalar and $\psi^i_X$ is a fermion field.
A CP transformation exchanges the operators 
\ba\label{CPoper}
\overline{\psi^i_{L}}\phi\psi^j_{R}\leftrightarrow
\overline{\psi^j_{R}}\phi^\dagger\psi^i_{L},
\ea 
but leaves their coefficients, $Y_{ij}$ and $Y_{ij}^*$, unchanged, since CP is a linear unitary non-anomalous transformation. This means 
that CP is conserved if 
\ba 
Y_{ij}=Y_{ij}^*\,.
\ea
This is, however, not a basis independent statement.
Since physical observables do no depend on a specific basis choice it is enough to 
find a basis in which the above relation holds.\footnote{It is easy to show that in this example, in fact, CP is not violated for any number of generations.} 

Sometimes the brute force way is tedious and might be quite complicated.
A more systematic approach would be to identify a phase reparamaterization invariant
or basis independentd quantity, that vanishes in the CP conserving limit. 
As discovered in~\cite{Jarlskog}, for the SM case one can define the following quantity  
\ba\label{JarCon}
\CSM =\det[Y_D Y_D^\dagger,Y_U Y_U^\dagger]\,,
\ea
and the SM is CP violating if and only if
\ba\label{JarCon}
\Im \left(\CSM \right)\neq0.
\ea
It is trivial to prove that only if the number of generations is three or more then CP is violated.
Hence, within the SM, where CP is broken explicitly in the flavor sector, any CP violating process has to involve all the three generations. This is a strong requirement
which leads to several sharp predictions.
Furthermore, all the CPV observables are correlated since they are all proportional to a single CP violating parameter, $\delKM$.
Finally, it is worth mentioning that CP violating observables are related to interference between different processes and hence are
measurements of amplitude ratios. Thus, in various known cases they turn out to be cleaner and easier to interpret 
theoretically. 

\subsection{The flavor puzzle}

Now that we have precisely identified the SM physical flavor parameters it is interesting to ask what is their experimental value (using $\rm \overline{MS}$)~\cite{PDG}:
\ba\label{flavorpara}
m_u&=& 1.5..3.3 \MeV \,, \ m_d=3.5..6.0 \MeV\,, \ m_s = 150^{+30}_{-40}\MeV\,, \nn \\
m_c&=&1.3\GeV\,, \ m_b=4.2 \GeV\,, \ m_t = 170\GeV\,, \nn \\
\VKM_{ud}&=&0.97\,, \ \VKM_{us}= 0.23\,, \ \VKM_{ub} = 3.9\times 10^{-3}\,, \nn \\
\VKM_{cd}&=&0.23 \,, \ \VKM_{cs}=1.0 \,, \ \VKM_{cb} = 41\times 10^{-3}\,, \nn \\
\VKM_{td}&=&8.1\times 10^{-3}\,, \ \VKM_{ts}=39\times 10^{-3} \,, \ \VKM_{tb} = 1\,, \  \delKM=77^o\,,
\ea
where $\VKM_{ij}$ corresponds to the magnitude of the $ij$ entry in the CKM matrix, $\delta_{\rm KM}$ is the CKM phase,
only uncertainties bigger than~10\% are shown, numbers are shown
to a 2-digit precision and the $\VKM_{ti}$ entries involve indirect information
[a detailed description and Refs. can be found in the PDG~\cite{PDG}] .

Inspecting the actual numerical values for the flavor parameters, given in Eq. (\ref{flavorpara})
shows a peculiar structure. Most of the parameters, apart from the top mass and the CKM phase,
are small and hierarchical.
The amount of hierarchy in the flavor sector can be characterized by looking at two different classes of observables:
\bi
\item Hierarchies between the masses, which are not related to flavor converting processes - as a measure of these hierarchies we can
just estimate what is the size of the product of the Yukawa coupling square differences 
\ba
{\left(m_t^2-m_c^2\right) \left(m_t^2-m_u^2\right)  \left(m_c^2-m_u^2\right) \left(m_b^2-m_s^2\right) \left(m_b^2-m_d^2\right)   \left(m_s^2-m_d^2\right) \over v^{12}}=\Ord{10^{-19}}\,.
\ea
\item Hierarchies in the mixing which mediate flavor conversion, this is related to the tiny misalignment between the up and down Yukawas -
one can quantify this effect in a basis independent fashion as follows.
 A CP violating quantity, associated with $\VKM$, that is independent 
of  parametrization \cite{Jarlskog}, $J_{\rm KM}$, is 
defined through 
\ba\label{defJ}
\hspace*{-.405cm}\Im\big[\VKM_{ij}\VKM_{kl} \big(\VKM_{il}\big)^*\big(\VKM_{kj}\big)^*\big]=
J_{\rm KM}\sum_{m,n=1}^3\epsilon_{ikm}\epsilon_{jln}
=c_{12}c_{23}c_{13}^2s_{12}s_{23}s_{13}\sin\delKM\simeq \lambda^6 
A^2\eta=\Ord{10^{-5}},
\ea
where $i,j,k,l=1,2,3$. We see that even though $\delKM$ is of order unity the resulting CP violating parameter
is small since it is "screened" by small mixing angles. If any of the mixing angles is a multiple of $\pi/2$ then
the SM Lagrangian becomes real.
Another, explicit way to see that $ Y_U$ and $Y_D$ are quasi aligned is via the Wolfenstein 
parametrization of the CKM matrix, where the four mixing parameters are $(\lambda,A,\rho,\eta)$ 
with $\lambda=|V_{us}|=0.23$ playing the role of an expansion parameter~\cite{Wolfenstein}:
\ba
\label{WCKM}
\VKM=\begin{pmatrix} 1-{\lambda^2\over2}&\lambda&A\lambda^3(\rho-i\eta)\cr
-\lambda&1-{\lambda^2\over2}&A\lambda^2\cr
A\lambda^3(1-\rho-i\eta)&-A\lambda^2&1\cr\end{pmatrix}+{\cal O}(\lambda^4).
\ea
Basically, to zeroth order, the CKM matrix is just a unit matrix\,!
\ei
As we shall discuss further below, both kind of hierarchies described in the bullets lead to suppression of CPV. Thus, a nice way to quantify the 
amount of hierarchies both in masses and mixing angles is to compute the value of the reparameterization invariant measure
of CPV introduced in Eq. (\ref{JarCon})
\ba\label{JarConSM}
\CSM =J_{\rm KM} \times {\left(m_t^2-m_c^2\right) \left(m_t^2-m_u^2\right)  \left(m_c^2-m_u^2\right) \left(m_b^2-m_s^2\right) \left(m_b^2-m_d^2\right)   \left(m_s^2-m_d^2\right) \over v^{12}}=\Ord{10^{-23}}\,.
\ea
This tiny value of $\CSM$ that characterizes the flavor hierarchy in nature would be of order 10\% 
in theories where
$Y_{U,D}$ are generic order one complex matrices.
The smallness of $\CSM$ is something that many flavor models beyond the SM try to address.
Furthermore, SM extensions that have new sources of CPV tend not to have the SM built in CP screening mechanism. 
Thus, they give too large contributions to the various observables
that are sensitive to CP breaking. Therefore, these models are usually excluded by the data, which is consistent with the SM predictions.

\section{Spurion analysis of the SM flavor sector \& minimal flavor violation}\label{spurion}
In this part we shall try to be more systematic in understanding the way flavor is broken within the SM.
We shall develop a spurion, symmetry oriented description for the SM flavor structure and also
generalize it to NP models with similar flavor structure that goes under the name minimal flavor violation (MFV).

\subsection{Spurion understanding of the SM flavor breaking}\label{spurionsub}
\begin{figure}[h]   
  \begin{center}
    \includegraphics[width=0.5\textwidth]{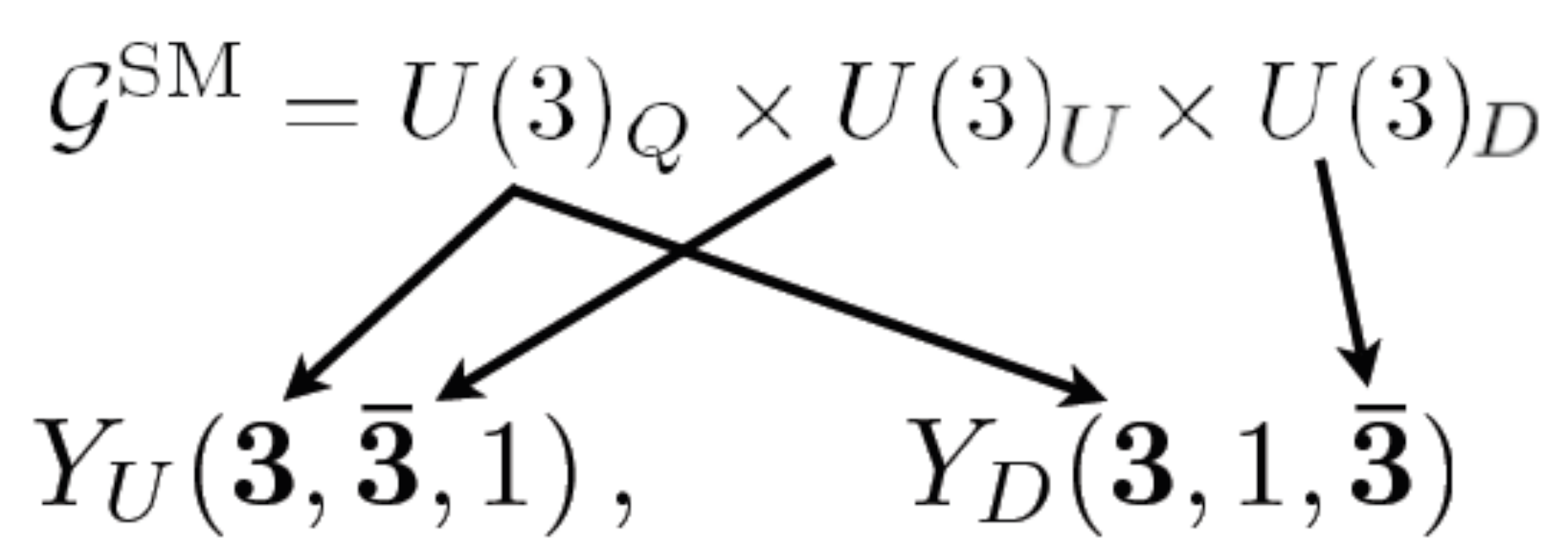}
    \caption{The SM  flavor symmetry  breaking by the Yukawa matrices.}
    \label{MFVbreaking}
  \end{center}
\end{figure}

\begin{figure}[h]   
  \begin{center}
    \includegraphics[width=0.75\textwidth]{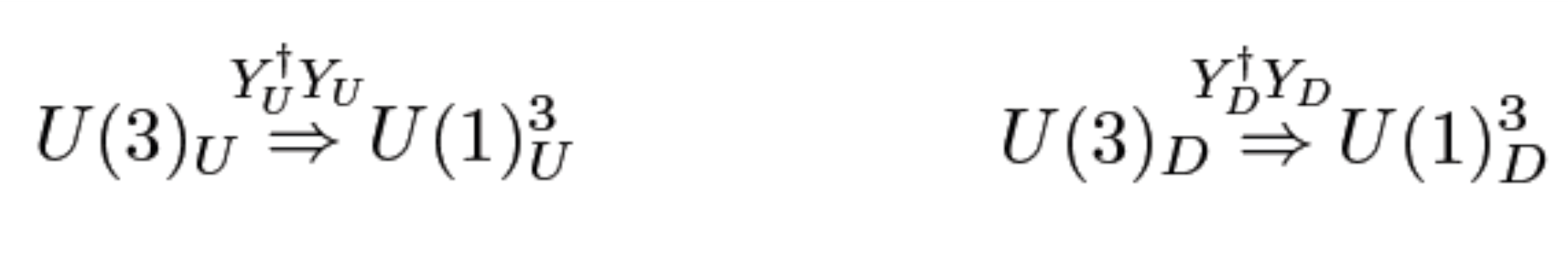}
    \caption{Breaking of the $U(3)_{U,D}$ groups by the Yukawa matrices which form an appropriate LH (RH) flavor group singlet (adjoint+singlet).}
    \label{RHbreaking}
  \end{center}
\end{figure}

It is clear that if we set the Yukawa couplings of the SM to zero we restore the full global flavor group,
$ \GSM= U(3)_Q\times U(3)_U\times U(3)_D\,.$
In order to be able to better understand the nature of flavor and CPV within the SM,  in the presence of the Yukawa terms, we can use a spurion analysis as follows.
Let us formally promote the Yukawa matrices to spurion fields, which transform under $\GSM$
in a manner that makes the SM invariant under the full flavor group (see \eg~\cite{MFVspurions} and Refs. therein).
From the flavor transformation given in Eqs. (\ref{Qtrans},\ref{Ytrans}) we can read the representation
of the various fields under $\GSM$ (see illustration in Fig. \ref{MFVbreaking})
\ba
{\rm Fields:}&&\ \ Q(\mathbf{3},1,1),U(1,\mathbf{3},1), D(1,1,\mathbf{ 3})\,; \nn \\
{\rm Spurions:}&&\ \ Y_U(\mathbf{3},\mathbf{\bar 3},1), Y_D(\mathbf{3},1,\mathbf{\bar 3})\,. 
\ea
The flavor group is broken by the "background" value of the spurions $Y_{U,D}$ which are bi-fundamentals
of $\GSM$. 
It is instructive to consider the breaking of the different flavor groups separately (since $Y_{U,D}$ are 
bi-fundamentals the breaking of quark doublet and singlet flavor groups are linked together, so this analysis only give partial information to be completed below).
Consider the quark singlet flavor group, $U(3)_U\times U(3)_D$, first.
We can construct a polynomial of the Yukawas with simple transformation properties under 
the flavor group.
For instance, consider the objects 
\ba
A_{U,D}\equiv Y_{U,D}^\dagger Y_{U,D}-{1\over3}\tr\left(Y_{U,D}^\dagger Y_{U,D}\right)\mathbf{I}_3\,.
\ea
Under the flavor group $A_{U,D}$ tranform as
\ba\label{AUD}
A_{U,D}\to V_{U,D} A_{U,D} V_{U,D}^\dagger\,.
\ea
 Thus, $A_{U,D}$ are adjoints of $U(3)_{U,D}$ and singlets of the rest of the flavor group 
 [while $\tr(Y_{U,D}^\dagger Y_{U,D})$ are flavor singlets].
 Via similarity transformation we can bring $A_{U,D}$ to a diagonal form, simultaneously.
 Thus, we learn that the background value of each of the Yukawa matrices separately breaks the $U(3)_{U,D}$
 down to a residual $U(1)^3_{U,D}$ group, as illustrated in Fig. \ref{RHbreaking}.

Let us now discuss the breaking of the LH flavor group.
We can, in principle, apply the same analysis for the LH flavor group, $U(3)_Q$,
via defining the adjoints (in this case we have two independent ones),
\ba\label{AQ}
A_{Q^u,Q^d}\equiv Y_{U,D} Y_{U,D}^\dagger-{1\over3}\tr\left(Y_{U,D} Y_{U,D}^\dagger\right)\mathbf{I}_3\,.
\ea
However, in this case the breaking is more involved since $A_{Q^{u,d}}$ are adjoint of the same flavor group.
This is a direct consequence of the $SU(2)$ weak gauge interaction which relates the two components of the $SU(2)$
doublets.
This actually motivates one to extend the global flavor group as follows.
If we switch off the electroweak interactions the SM global flavor group is actually enlarged to 
\ba
  \label{GSMweakless}
  \GSM_{\rm weakless}= U(6)_Q\times U(3)_U\times U(3)_D\,,
\ea
since now each $SU(2)$ doublet, $Q_i$ can be split into two independent flavors, $Q_i^{u,d}$
with identical $SU(3)\times U(1)$ gauge quantum numbers~\cite{weakless}.
This limit, however is not very illuminating since it does not allow for flavor violation at all. 
To make a progress it is instructive to distinguish the $W^3$ neutral current interactions from the $W^\pm$
charged current ones as follows:
The $W^3$ couplings are flavor universal, which, however, couple up and down
quarks separately. 
The $W^\pm$ couplings, $\gW$, link between the up and down LH quarks. In the presence of only $W^3$ couplings the residual flavor group is given by\footnote{To get to this limit, formally, one can think of a model where the Higgs field is an adjoint of $SU(2)$
and a singlet of color and hypercharge. In this case the Higgs VEV preserves a $U(1)$ gauge symmetry and the $W^3$
would therefore still remain massless. However, the $W^\pm$ will acquire masses of the order of the Higgs VEV and, therefore, charged current interactions would be suppressed.}
\ba
  \label{GSMW3}
  \GSME= U(3)_{Q^u}\times  U(3)_{Q^d}\times U(3)_U\times U(3)_D\,.
\ea 
In this limit, even in the presence of the Yukawa matrices, flavor conversion is forbidden since we have already seen explicitly that
only the charged currents link between different flavors [see \Eq{Lflavormass}].
It is, thus, evident that to formally characterize flavor violation we can extend the flavor group from $\GSM\to \GSME$
where now we break the quark doublets to their isospin components, $U_L,D_L$, and add another spurion, $\gW$ 
\ba\label{flavextent}
{\rm Fields:}&&\ \ U_L(\mathbf{3},1,1,1), D_L(1,\mathbf{3},1,1), U(1,1,\mathbf{3},1), D(1,1,1,\mathbf{ 3}) \nn \\
{\rm Spurions:}&&\ \ \gW(\mathbf{3},\mathbf{\bar 3},1,1), Y_U(\mathbf{3},1,\mathbf{\bar 3},1), Y_D(1,\mathbf{3},1,\mathbf{\bar 3}) \,.
\ea
Flavor breaking within the SM occurs only when $\GSME$ is fully broken via the Yukawa background values but
also due to the fact that $\gW$ has a background value. Unlike $Y_{U,D}$, $\gW$ is a special spurion in the sense 
that its eigen values are degenerate as required by the weak gauge symmetry.
Hence, it breaks the $U(3)_{Q^u}\times U(3)_{Q^d}$ down to a diagonal group which is nothing but $U(3)_Q$.
We can identify two bases where $\gW$ has an interesting background value: 
The weak interaction basis where the background value of $\gW$ is simply a unit matrix\footnote{Note that the interaction basis is not unique since $\gW$ is invariant under flavor transformation where $Q^u$ and $Q^d$ are rotated by the same amount, see more in the following.}
\ba\label{gWint}
\left(\gW\right)_{\rm int}\propto \mathbf{1}_3\,,
\ea
and the mass basis where (after removing all unphysical parameters) the background value of $\gW$ is the CKM matrix
\ba
\left(\gW\right)_{\rm mass}\propto \VKM\,.
\ea

\begin{figure}[h]   
  \begin{center}
    \includegraphics[width=.87\textwidth]{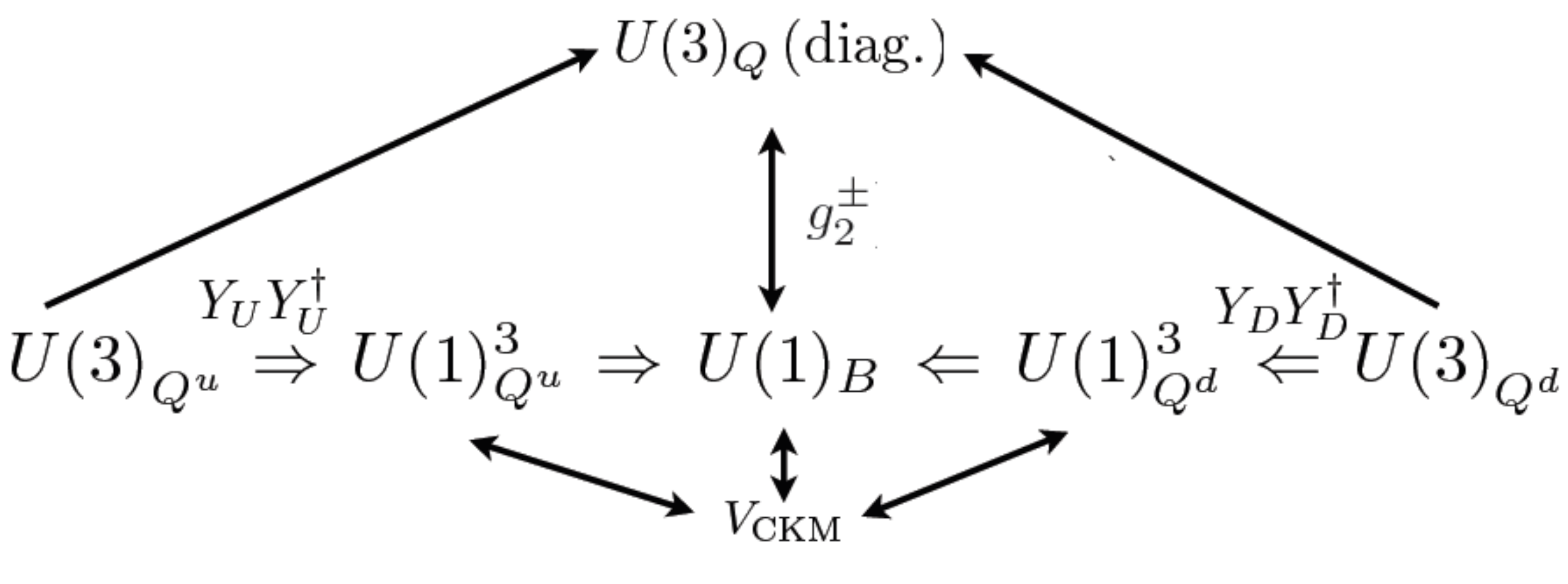}
    \caption{$U(3)_{Q_{u,d}}$ breaking by $A_{Q^u,Q^d}$ and $\gW$.}
    \label{LHbreaking}
  \end{center}
\end{figure}

\begin{figure}[h]   
  \begin{center}
    \includegraphics[width=.87\textwidth]{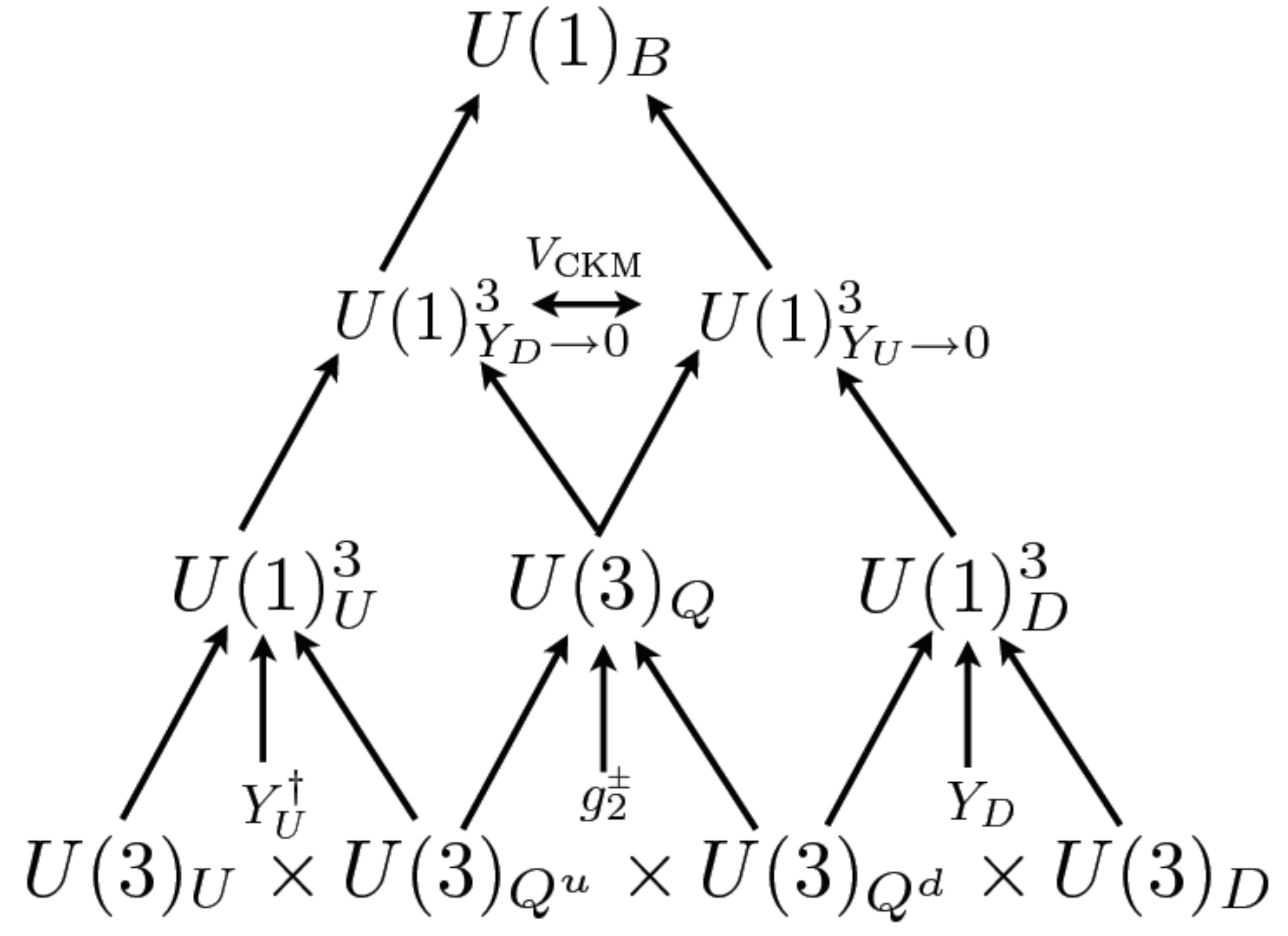}
    \caption{The schematic structure of the various ingredients that mediate flavor breaking within the SM.}
    \label{SMflavorbreaking}
  \end{center}
\end{figure}

Now we are at position to understand the way flavor conversion is obtained in the SM.
Three spurions must participate in the breaking, $Y_{U,D}$ and $\gW$.
Since $\gW$ is involved it is clear that generation transition has to involve LH charged current
interactions. These transitions are mediated by the spurion backgrounds, $A_{Q^u,Q^d}$ [see Eq. (\ref{AQ})]
which characterize the breaking of the individual LH flavor symmetries, 
\ba U(3)_{Q^u}\times U(3)_{Q^d}\to U(1)^3_{Q^u}\times U(1)^3_{Q^d}\,.\ea
Flavor conversion occurs because of the fact that in general we cannot diagonalize
simultaneously $A_{Q^u,Q^d}$ and $\gW$, where the misalignment between $A_{Q^u}$ and $A_{Q^d}$
is precisely characterized by the CKM matrix.
This is illustrated in Fig. \ref{LHbreaking},
where it is shown that the flavor breaking within the SM goes through collective breaking~\cite{GMFV} a term
often used in the context of little Higgs models (see \eg~\cite{LHRev} and Refs. therein).
We can now combine the LH and RH quark flavor symmetry breaking to obtain the complete picture of how flavor is broken within the SM.
As we saw the breaking of the quark singlet groups is rather trivial. It is, however, linked to the more involved LH flavor breaking  since the Yukawa matrices are bi-fundamentals -- The LH and RH flavor breaking are tied together.
The full breaking is illustrated in Fig.~\ref{SMflavorbreaking}.

\subsection{A comment on description of flavor conversion in physical processes}\label{MFV}
The above spurion structure allows us to describe SM flavor converting processes.
However, the reader might be confused since we have argued above that flavor converting processes 
must involve the three spurions, $A_{Q^{u,d}}$ and $\gW$.
It is well known, that the rates for charge current processes, which are described via conversion of 
down quark to an up one (and vise a versa), say like beta decay or $b\to u$ transitions are
only suppressed by the corresponding CKM entry, or $\gW$.
What happened to the dependence on $A_{Q^{u,d}}$?
The key point here is that in a typical flavor precision measurement the experimentalists 
produce mass eigenstate (for example a neutron or a $B$ meson), and thus the fields involved are
chosen to be in the mass basis. For instance a $b\to c$ process is characterized by 
producing a $B$ meson which decays into a charmed one. Hence, both $A_{Q^{u,d}}$, $A_{Q^{u}}$
participate, being forced to be diagonal, but in a nonlinear way ({\it i.e} strictly speaking this transition cannot be described in a basis independent fashion by some simple insertion of powers of $A_{Q^{u}}$ and $A_{Q^{d}}$).
 Physically we can characterize it by writing an operator 
\ba{\cal O}_{b\to c}=\bar c_{\rm mass} \left(\gW\right)^{cb}_{\rm mass} b_{\rm mass}\,
\ea
where both the $b_{\rm mass}$ and $c_{\rm mass}$ quarks are mass eigenstate.
Note that this is consistent with the transformation rules for the extended gauge group, $\GSME$ given in Eqs.~(\ref{GSMW3},\ref{flavextent}), where the fields involved belong to different representations of the extended flavor group.

The situation is different when flavor changing neutral current processes are considered.
In such a case a typical measurement involves mass eigenstate quarks belonging to the same representation of $\GSME$.
For example, processes that mediate $B^0_d-\bar B^0_d$ oscillation due to the tiny mass difference $\Delta m_d$ between the two mass eigenstates (which was measured for the first time by the ARGUS experiement~\cite{ARGUS}) are described via the following operator, omitting the spurion structure for simplicity,
\ba
{\cal O}_{\Delta m_d}=\left(\bar b_{\rm mass}\, d_{\rm mass}\right)^2\,.
\ea
Obviously, this operator cannot be generated by SM processes since it is violates the $\GSME$ symmetry. 
Since it involves flavor conversion (it violates $b$ number by two units, hence denoted as $\Delta b=2$ and belongs to $\Delta F=2$ class of FCNC processes)
it has to have some power of $\gW$. A single power of $\gW$ connects
LH down quark to a LH up one, so the leading contribution should go like $\bar D_L^i \left(\gW\right)^{ik} \left(\gW^*\right)^{kj} D_L^j$ ($i,k,j=1..3$) which implies, as expected that this process is mediated at least via one loop.
This would not work as well since we can always rotate the down quark fields into the mass basis and simultaneously rotate also the up type quarks (away from their mass basis)
so that  $\gW\propto \mathbf{1}_3$. These manipulations define the interaction basis which is not unique [see \Eq{gWint}].
Therefore, the leading flavor invariant spurion that mediates FCNC transition would have to involve the up type Yukawa spurion as well.
A naive guess would be
\ba\label{dmd}
{\cal O}_{\Delta m_d}&\propto&\left[\bar b_{\rm mass} \left(\gW\right)_{\rm mass} ^{bk} \left(A_{Q^u}\right)_{kl} \left(\gW^*\right)_{\rm mass} ^{ld} d_{\rm mass}\right]^2 \nn \\
&\sim& \left[\bar b_{\rm mass}  \left(m_t^2 \left(\VKM\right)_{tb} \left(\VKM\right)^*_{td}+m_c^2 \left(\VKM\right)_{cb} \left(\VKM\right)^*_{cd}\right) d_{\rm mass}\right]^2\,,
\ea
where it is understood that $\left(A_{Q^u}\right)_{kl}$ is evaluated in the down quark mass basis (obviously tiny corrections of order  $m_u^2$ are neglected in the above).
This expression captures the right flavor structure and is correct for large class of SM extensions.
However, it is actually incorrect in the SM case. The reason is that within the SM
the flavor symmetries are badly broken by the large top quark mass~\cite{GMFV}.
The SM corresponding amplitude consist of a rather non-trivial, non-linear function of $A_{Q^u}$ instead of the above naive
expression~(see \eg \cite{BBL} and Refs. therein), which assumes only the simplest polynomial dependence of the spurions.  
The SM amplitude for $\Delta m_d$ is described via a box diagram and two out of the four power of masses
are cancelled, since they appear in the propagators.

\subsection{The SM approximate symmetry structure}\label{MFV}

 In the above we have considered the most general breaking pattern. However,
 as we have discussed the essence of the flavor puzzle is the large hierarchies in the quark masses, the eigen values of
 $Y_{U,D}$ and their approximate allignment.
 Going back to $A_{Q^{u,d}}$ [defined in Eqs. (\ref{AUD},\ref{AQ})], the spurions that mediate the SM flavor conversions, we can write them as
 \ba\label{approx}
A_{U,D}=\diag\left(0,0,y_{t,b}^2\right)-{y_{t,b}^2\over3}\mathbf{1}_3+\Ord{m_{c,s}^2\over m_{t,b}^2}\,, \ \ \
A_{Q^{u,d}}=\diag\left(0,0,y_{t,b}^2\right)-{y_{t,b}^2\over3}\mathbf{1}_3+\Ord{m_{c,s}^2\over m_{t,b}^2}+\Ord{\lambda^2}\,,
\ea
where in the above we took advantage of the fact that ${m_{c,s}^2\over m_{t,b}^2},\lambda^2=\Ord{10^{-5,-4,-2}}$
are small. 
The large hierarchies in the quark masses is translated to an approximate residual RH $U(2)_U\times U(2)_D$ flavor group
which implies that RH currents which involve light quarks are very small.

We have so far only briefly discussed the role of flavor changing neutral currents (FCNCs).
In the above we have argued, both based on an explicit calculation and in terms of a spurion analysis, that
at tree level there are no flavor violating neutral currents, since they must be mediated through the $W^\pm$ couplings
or $\gW$.
In fact, this situation, which is nothing but the celebrated GIM mechanism~\cite{GIM}, goes beyond the SM
to all models where all LH quarks are $SU(2)$ doublets
and all RH ones are singlets. The $Z$-boson might
have flavor changing couplings in models where this is not the case. 

Can we guess what is the leading spurion structure that induces FCNC within the SM, say which mediates 
the $b\to d\nu\bar \nu$ decay process, via an operator $\Obdn$? 
The process changes $b$ quark number by one unit (belongs to $\Delta F=1$ class of FCNC transitions).
It clearly has to contain down type, LH, quark fields (let us ignore the lepton current which is flavor-trivial, for effects related to
neutrino masses and lepton number breaking in this class of models see \eg~\cite{KLn}). 
Therefore, using the argument presented when discussing $\Delta m_d$ [see \Eq{dmd}], the leading flavor invariant spurion that mediate FCNC would have to involve the up type Yukawa spurion as well
\ba
\Obdn \propto \bar D_L^i \gW_{ik} \left(A_{Q^u}\right)_{kl} \gW^*_{lj} D_L^j \times\bar \nu\nu\,.
\ea
The above considerations demonstrate how the GIM mechanism removes the SM divergencies from various one loop FCNC processes.
These, naively, are expected to be log divergent. The reason is that the insertion of $A_{Q^u}$
is translated to quark mass difference insertion. It means that the relevant one loop diagram
has to be proportional to $m_i^2-m_j^2$ ($i\neq j$). Thus, the superficial degree of divergency is lowered
by two units which renders the amplitude finite.\footnote{For simplicity we only consider cases with hard GIM where the dependence on mass differences is polynomial. There is a large class of amplitudes, for example processes that are mediated via penguin diagrams with gluon or photon lines, where the quark mass dependence is more complicated and may involve logarithms. The suppression of the corresponding amplitudes goes under the name soft GIM~\cite{BBL}. }
Furthermore, as explained above [see also~\Eq{tdom}] we can use the fact that the top contribution dominates the flavor violation to
simplify the form of $\Obdn$
\ba\label{estimate}
\Obdn \sim {g_2^4\over 16\pi^2 M_W^2}\,\bar b_L  \left(\VKM\right)_{tb}\left(\VKM\right)^*_{td} d_L\times \bar \nu \nu\,.
\ea
where we have added a one loop suppression factor and expected weak scale suppression.
This rough estimation actually reproduces the SM result up to a factor of about 1.5 (see \eg~\cite{BBL,BurasMFV}).
We, thus, find that down quark FCNC amplitudes are expected to be highly suppressed due to the smallness of
the top off-diagonal entries of the CKM entries.
Parameterically we find the following suppression factor for transition between the $i$th and $j$th generations:
\ba\label{fsup}
b\to s &\propto&  \big|\left(\VKM\right)_{tb} \left(\VKM\right)_{ts}\big| \sim \lambda^2\,, \nn\\
b\to d &\propto & \big|\left(\VKM\right)_{tb} \left(\VKM\right)_{td}\big| \sim \lambda^3\,, \nn \\
s\to d &\propto&   \big|\left(\VKM\right)_{td} \left(\VKM\right)_{ts}\big|  \sim \lambda^5\,, 
\ea 
where for the $\Delta F=2$ case one needs to simply square the parameteric suppression factors.
This simple exercise illustrates how powerful is the SM FCNC suppression mechanism.
The gist of it is that the rate of SM FCNC processes is small since they occur at one loop, and more importantly
due to the fact that they are suppressed by the top CKM off-diagonal entries, which 
are very small.
Furthermore, since 
\ba \label{tdom}
\left|\VKM_{ts,td}\right|\gg {m_{c,u}^2\over m_{t}^2}\,,
\ea
in most cases the dominant flavor conversion effects
are expected to be mediated via the top Yukawa coupling.\footnote{This definitely is correct for CP violating processes or any ones which involve the third generation quarks. It also, generically, holds for new physics MFV models.
Within the SM, for CP conserving processes which involve only the first two generations one
can find exceptions, for instance when considering the kaon and $D$ meson mass differences,  $\Delta m_{D,K}$.} 

We can now understand how the SM uniqueness related to suppression of flavor converting processes arises:
\bi
\item RH currents for light quarks are suppressed due to their small Yukawa couplings (them being light). 
\item Flavor transition occurs to leading order only via LH charged current interactions.
\item To leading order, flavor conversion is only due to the large top Yukawa coupling.
\ei

\subsection{Minimal flavor violation}\label{MFV}

So far we have focused only on the SM flavor structure and developed a spurion description 
of the SM flavor breaking.
We can, however, extend our above analysis to include also an important class 
of SM extensions, denoted as minimal flavor violation (MFV)~\cite{MFV} which includes, among others, various extended Higgs models~\cite{MFV,2HDM}, supersymmetric models~\cite{GMSB,AMSB}
and under some assumptions warped extra dimension models~\cite{GMFV,shining} (for reviews see \eg~\cite{BurasMFV} and Refs. therein).
As we shall see, the models which belong to the MFV class enjoy much of the protection against large flavor violation
that we have found to exist in the SM case and therefore tend to be consistent with current flavor precision measurements.

The basic idea can be described in the language of effective field theory (EFT) without the need of referring 
to a specific framework.
MFV models can have a very different microscopical dynamics, however, by definition they all have a common 
origin of flavor breaking, the SM Yukawa matrices.
After integrating out the NP degrees of freedom we expect to obtain a low energy EFT which involves only the SM
fields and bunch of higher dimension, Lorentz and gauge invariant, operators suppressed by the NP scale $\LMFV$.
Since flavor is broken only via the SM Yukawas then we can study the most general 
flavor breaking of the MFV framework by the simple following prescription:  We  should construct the most general set of higher dimensional operators, which 
in addition of being Lorentz and gauge invariant, they are required also to be flavor invariant, using the spurion 
analysis that we have introduced in the previous part.

If we are interested in SM processes where the typical energy scale is much smaller than $\LMFV$ and the NP is not strongly coupled then we expect that the dominant non-SM flavor violation would arise from the lowest order higher dimension operators.
For processes involving quark fields, the leading operators are of dimension six. 
For instance, we expect that the leading flavor violating operators that mediate  $\Delta F=2$ processes would involve only LH fields,\footnote{Note that in the presence of NP, we do not generally expect that $\gW$ would be the only object that mediates the breaking of $U(3)_{Q^u}\times U(3)_{Q^d}$, hence there is no advantage in using $\GSME$ representations in this case, nor $\gW$ as a spurion.} 
\ba
\LMFVF={1\over (\LMFV)^2}\left[\bar Q_i\left( a_u A_{Q^u} +a_d A_{Q^d} \right)Q_j\right]^2+\dots
\ea
where in the above we have written, for simplicity, the leading polynomial of $Y_{U,D}$ which mediates flavor conversion.\footnote{Since the top Yukawa is of order unity and possibly also the bottom one this might not be a justified assumption.
Generically, one would expect to get a generic polynomial of the Yukawas instead of just a quadratic term as presented here.
The general case can be dealt with by resumming the contributions from the large eigenvalue via non-linear sigma model 
techniques, which allow one to separate the large and small terms in the Yukawa matrices~\cite{GMFV}.
This leads to somewhat richer structure than discussed here. For simplicity we shall only focus on the linear MFV case where only the leading terms in the polynomial expansion are considered.}
Let us, for instance, focus on flavor violation in the down sector which is most severely constrained. 
We want to estimate what is the size of flavor violation which is mediated by $\LMFVF$.
The experimental information is obtained by looking at the dynamics (masses, mass differences, decay, time evolution etc...)
of down type mesons hence we can just look at the form that $\LMFVF$ takes in the down quark mass basis.
By definition $A_{Q^d}$ is diagonal and does not mediate flavor violation, however, in the down type basis,
$A_{Q^u}$ is not diagonal and is given by
\ba
\left(A_{Q^u}\right)_{\rm down}=\VKM \diag\left(0,0,y_t^2\right)\left(\VKM\right)^\dagger-{y_t^2\over3}\mathbf{1}_3+\Ord{m_c^2\over m_t^2}\sim y_t^2  \left(\VKM\right)_{ti}\left(\VKM\right)^*_{tj} \,,
\ea
where in the above we took advantage of the approximate $U(2)$ symmetry limit discussed in the previous subsection.
As expected, we find that within the MFV framework FCNC processes are suppressed by roughly the same
amount as the SM processes, and therefore are typically, at least to leading order, consistent with present data.
This need not to be the case when new flavor diagonal CP violating sources are allowed~\cite{GMFV,CPdiag}.

\begin{figure}[h]   
  \begin{center}
   \hspace*{-.38cm} \includegraphics[width=.5\textwidth]{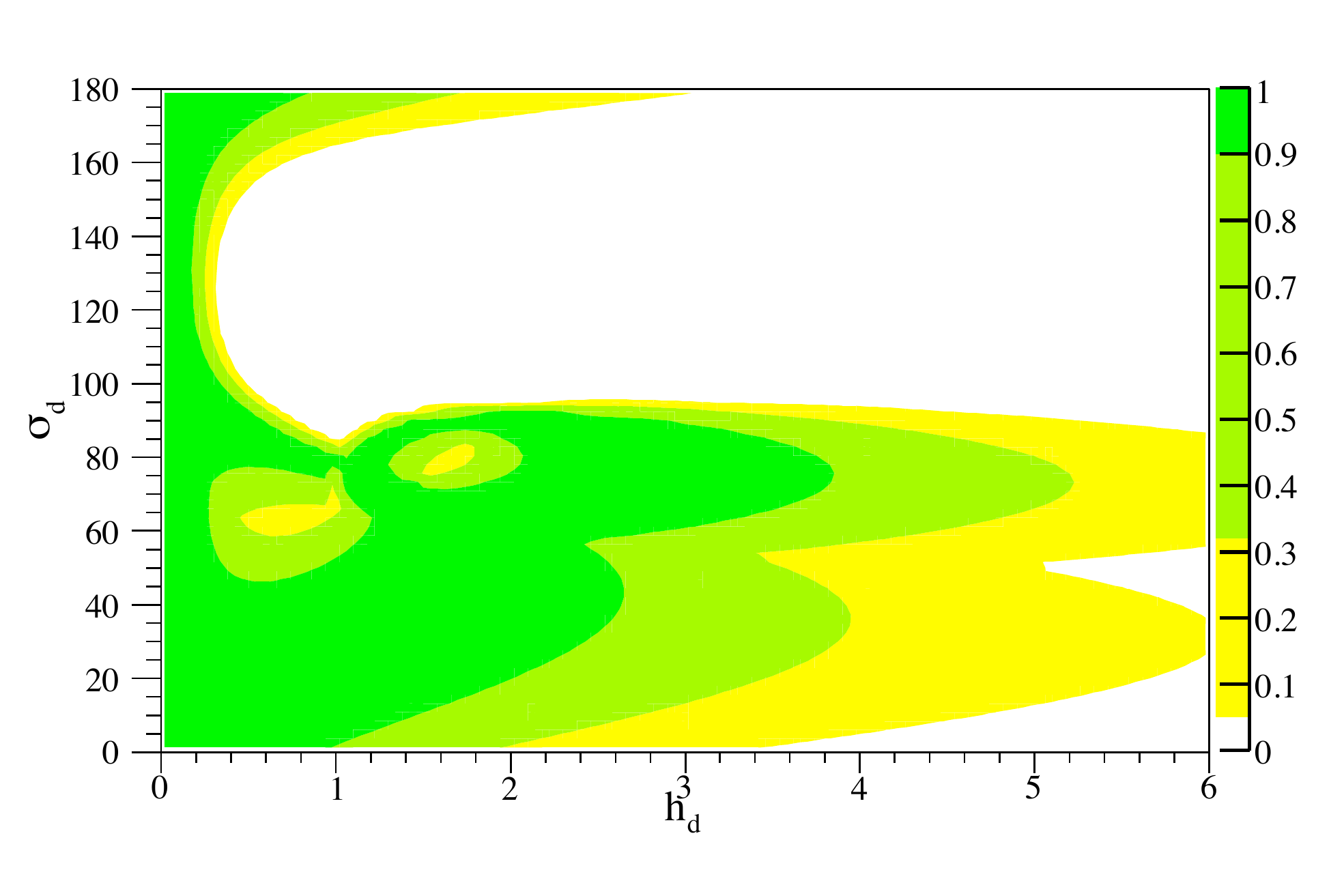}  \includegraphics[width=.5\textwidth]{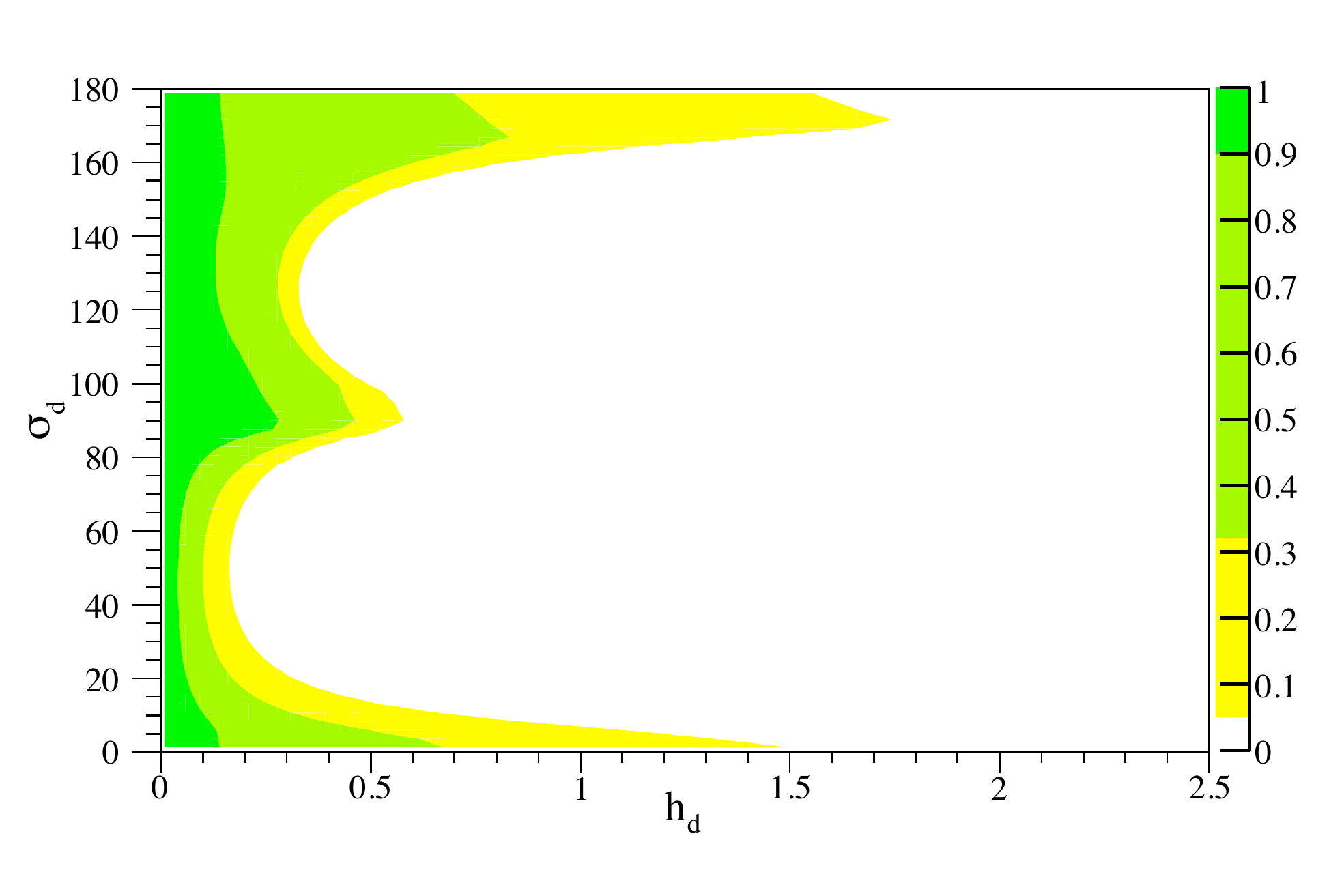}
    \caption{On the left (right) the allowed range for the new physics contributions to $\Delta F=2$ processes in the $B^0_d$ system. The constraints are shown in the $h_d-\sigma_d$ plan prior to (after) 2004~\cite{NMFV}. The color scale represents the confidence level (1 means only unknown theoretical uncertainties included). Measurements of SM, tree-level, CPV observables, which are unlikely to be affected by new physics, led to an order of magnitude improvement in the constraints of non-SM contributions.
    It is established that the SM contributions are the dominant ones.}
    \label{hdsd}
  \end{center}
\end{figure}

\subsection{Beyond MFV \& the 2004 "revolution"}

It is interesting to note that only fairly recently has the data begun to disfavor models with only LH currents, 
but with new sources of flavor and CPV~\cite{Ligeti,NMFV}, 
 characterized by a CKM-like suppression~\cite{Davidson:2007si,aps}.
In fact this is precisely the way that one can test the success of Kobayashi-Maskawa mechanism for flavor 
and CP violation~\cite{NMFV,Buras:2009us,UTFit,CKMFitter,test}. 
Below we focus on NP in $\Delta F=2$ processes which are clean to interpret theoretically.  In addition, for simplicity, we only focus on the $B_d$ system
where the improvement in constraining new data was particularly dramatic.
The NP contributions to $B_d^0$ mixing can be expressed in terms of
two parameters, $h_d$ and $\sigma_d$ defined by 
\ba
M_{12}^d = (1+h_d
e^{2i\sigma_d}) M_{12}^{d,{\rm SM}}\,,
\ea 
where $M_{12}^{d,{\rm SM}}$ is the
dispersive part of the $B^0_d-\bar B^0_d$ mixing amplitude in the SM. 

\begin{figure}[h]   
  \begin{center}
   \includegraphics[width=.75\textwidth]{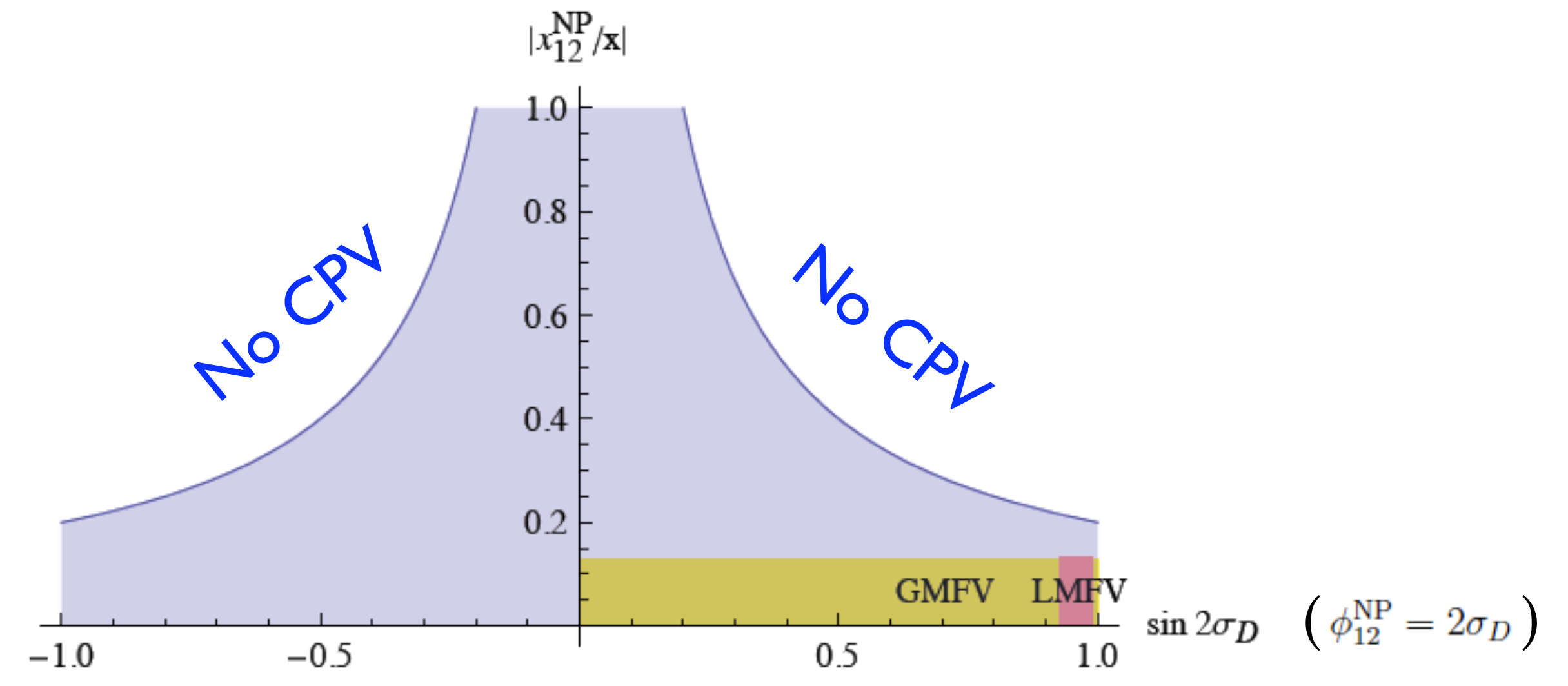} 
    \caption{The allowed region, shown in grey, in the
$x_{12}^{\rm NP}/x_{12}-\sin\phi_{12}^{\rm NP}$ plane.
The pink and yellow  regions correspond to the ranges predicted by,
respectively, the linear MFV and general MFV classes of models~\cite{Gedalia:2009kh}.
}
    \label{hDsD_Dsystem}
  \end{center}
\end{figure}

To constrain deviations from the SM in these processes one can use 
measurements which are directly proportional to $M_{12}^d$ (magnitude and phase).
The relevant observables in this case are the neutral $B^0_d$ mass difference, $\Delta m_d$ and
the CPV in decay with and without mixing in $B^0_d\to \psi K$, $S_{\psi K}$.
These processes are characterized by hard GIM suppression and proceed, within the SM, via one loop 
[see Eqs. (\ref{estimate},\ref{fsup})], and in the presence of new physics they can be written as (see \eg~\cite{Nirrev1}):
\ba\label{par}
\Delta m_d &=& \Delta m_d^{\rm SM}\,
  \big|1+h_d e^{2i\sigma_d}\big| , \nn\\
S_{\psi K} &=& \sin \big[2\beta + \arg\big(1+h_d e^{2i\sigma_d}\big)\big]\,.
\ea
The fact that the SM contribution~\Eq{fsup} to these processes involve the CKM elements which are not measured directly
prevents one from independently constraining the NP contributions.
Indeed, prior to 2004, the experimental data yield the following rather weak constraints on $h_d$ and $\sigma_d$~\cite{NMFV},
\ba
 h_d\lesssim6 \ \ \ {\rm and} \ \ \  0\lesssim2\sigma_d \lesssim \pi\,.
\ea
The situation was dramatically improved when BaBar and Belle experiments managed to 
measure CPV processes which, within the SM, are mediated via tree level amplitudes.
The information extracted from these CP
asymmetries in $B^\pm\to DK^\pm$ and $B\to\rho\rho$ is probably hardly affected by
new physics. The most recent bounds (ignoring $2\sigma$ anomaly in $B\to \tau \nu$) are~\cite{CKMfitter09}
  \ba
 h_d\lesssim0.3 \ \ \ {\rm and} \ \ \  \pi\lesssim2\sigma_d \lesssim 2\pi\,,
\ea
which is very similar to the bound obtained just after 2004 when the new measurements became public.
Fig.~\ref{hdsd} shows the allowed range in the $h_d-\sigma_d$ plan before and after 2004 taken from~\cite{NMFV}.
Similar but less dramatic progress was made in the kaon and $B_s$ systems.
Furthermore, a recent progress in measurements of CPV in $D^0-\bar D^0$ mixing led
to an important improvement on the new physics constraints.
However, in this case the SM contributions are unknown~\cite{Dlong} and the only robust 
SM prediction is the absence of CPV~\cite{DCPV}.
The three relevant physical quantities
related to the mixing can be defined as
\ba\label{thepar}
y_{12}\equiv|\Gamma_{12}|/\Gamma,\qquad
x_{12}\equiv2|M_{12}|/\Gamma,\qquad
\phi_{12}\equiv\arg(M_{12}/\Gamma_{12})\,,
\ea
where $M_{12},\Gamma_{12}$ are the total
dispersive and absorptive part of the $D^0-\bar D^0$ amplitude respectively.
In Fig.~\ref{hDsD_Dsystem} we show (in grey) the allowed region in the
$x_{12}^{\rm NP}/x-\sin\phi_{12}^{\rm NP}$ plane. $x_{12}^{\rm NP}$ corresponds to the new physics contributions and
$x\equiv\frac{m_2-m_1}{\Gamma}$, where $m_i,\Gamma$ being the neutral $D$ meson mass eigenstates and average  width respectively. 
The pink and yellow  regions correspond to the ranges predicted by,
respectively, the linear MFV and general MFV classes of models~\cite{Gedalia:2009kh}.
We see that the absence of observed CP violation removes a sizable fraction of the possible new physics parameter space, in spite of the fact that the magnitude of the SM contributions cannot be computed!

\section{Conclusions}
In these rather laconic and far from being inclusive set of lectures, we have tried to 
develop a basic understanding of the SM flavor structure and some of its extensions.
The idea is to provide the readers with 
simple symmetry oriented principles, to understand the way flavor violation is mediated within the standard model (SM).
 The hope is that the methods described above would allow non-experts to understand
the unique SM flavor structure, and the power counting for suppression of various flavor changing processes.
Furthermore, understanding the SM mechanism for suppressing flavor changing neutral currents
also allows one to quickly estimate which models beyond the SM are likely to be excluded by current measurements.
In addition the analysis presented might even help the reader to identify viable models and directions for future tests.


\begin{acknowledgements}
The author thanks the organizer of the ISSCSMB '08 for the successful school and
great hospitality.
The author also thanks Oram Gedalia for comments on the manuscript.
 The work of GP is supported by the Israel Science Foundation (grant \#1087/09) and the Peter \&
Patricia Gruber Award.
\end{acknowledgements}

\end{document}